\pdfoutput=1
\documentclass[12pt,a4paper]{article}
\usepackage{ifthen} 
\newboolean{pdflatex}
\setboolean{pdflatex}{true} 

\newboolean{articletitles}
\setboolean{articletitles}{true} 

\newboolean{uprightparticles}
\setboolean{uprightparticles}{false} 
\def\paperauthors{LHCb collaboration}
\def\paperasciititle{Measurement of psi(2S) to J/psi cross-section ratio as function of multiplicity in pPb collisions at center of mass energy 8.16 TeV} 
\def\papertitle{Measurement of \psitwos to \jpsi cross-section ratio as function of multiplicity in $p$Pb collisions at $\sqsnn=8.16$ TeV}
\def\paperkeywords{{High Energy Physics}, {LHCb}} 
\def\papercopyright{\the\year\ CERN for the benefit of the LHCb collaboration} 
\def\paperlicence{CC BY 4.0 licence}
\def\paperlicenceurl{https://creativecommons.org/licenses/by/4.0/}

\usepackage[top=1in, bottom=1.25in, left=1in, right=1in]{geometry}

\usepackage{microtype}
\usepackage{lineno}  
\usepackage{xspace} 
\usepackage{caption} 

\usepackage{graphicx}  
\usepackage{color}
\usepackage{colortbl}
\graphicspath{{./figs/}} 

\usepackage{amsmath} 
\usepackage{amssymb}
\usepackage{amsfonts}
\usepackage{upgreek} 

\newcommand*\patchAmsMathEnvironmentForLineno[1]{%
\expandafter\let\csname old#1\expandafter\endcsname\csname #1\endcsname
\expandafter\let\csname oldend#1\expandafter\endcsname\csname
end#1\endcsname
 \renewenvironment{#1}%
   {\linenomath\csname old#1\endcsname}%
   {\csname oldend#1\endcsname\endlinenomath}%
}
\newcommand*\patchBothAmsMathEnvironmentsForLineno[1]{%
  \patchAmsMathEnvironmentForLineno{#1}%
  \patchAmsMathEnvironmentForLineno{#1*}%
}
\AtBeginDocument{%
\patchBothAmsMathEnvironmentsForLineno{equation}%
\patchBothAmsMathEnvironmentsForLineno{align}%
\patchBothAmsMathEnvironmentsForLineno{flalign}%
\patchBothAmsMathEnvironmentsForLineno{alignat}%
\patchBothAmsMathEnvironmentsForLineno{gather}%
\patchBothAmsMathEnvironmentsForLineno{multline}%
\patchBothAmsMathEnvironmentsForLineno{eqnarray}%
}

\usepackage{hyperxmp}

\usepackage[pdftex,
            pdfauthor={\paperauthors},
            pdftitle={\paperasciititle},
            pdfkeywords={\paperkeywords},
            pdfcopyright={Copyright (C) \papercopyright},
            pdflicenseurl={\paperlicenceurl}]{hyperref}

\usepackage[colorinlistoftodos,textsize=scriptsize]{todonotes}

\usepackage[bottom,flushmargin,hang,multiple]{footmisc}

\usepackage[all]{hypcap} 

\usepackage{xspace}
\usepackage{upgreek}

\def\lhcb   {\mbox{LHCb}\xspace}

\def\alice  {\mbox{ALICE}\xspace}

\def\lhc    {\mbox{LHC}\xspace}

\def\velo   {VELO\xspace}

\def\MagUp {\mbox{\em Mag\kern -0.05em Up}\xspace}

\ifthenelse{\boolean{uprightparticles}}%
{

 \def\Ppsi        {\ensuremath{\uppsi}\xspace}

 \def\PDelta      {\ensuremath{\Delta}\xspace}
 \def\PXi         {\ensuremath{\Xi}\xspace}
 \def\PLambda     {\ensuremath{\Lambda}\xspace}
 \def\PSigma      {\ensuremath{\Sigma}\xspace}
 \def\POmega      {\ensuremath{\Omega}\xspace}
 \def\PUpsilon    {\ensuremath{\Upsilon}\xspace}
 \let\oldPi\Pi
 \def\PPi         {\ensuremath{\oldPi}\xspace}

 \def\PB      {\ensuremath{\mathrm{B}}\xspace}
 \def\PD      {\ensuremath{\mathrm{D}}\xspace}
 \def\PJ      {\ensuremath{\mathrm{J}}\xspace}
 \def\PK      {\ensuremath{\mathrm{K}}\xspace}
 \def\Pb      {\ensuremath{\mathrm{b}}\xspace}
 \def\Pc      {\ensuremath{\mathrm{c}}\xspace}

 \def\Pp      {\ensuremath{\mathrm{p}}\xspace}

 \def\Ps      {\ensuremath{\mathrm{s}}\xspace}

 \def\thebaroffset{0.0em}
}
{

 \def\Ppsi        {\ensuremath{\psi}\xspace}
 
 \mathchardef\PDelta="7101
 \mathchardef\PXi="7104
 \mathchardef\PLambda="7103
 \mathchardef\PSigma="7106
 \mathchardef\POmega="710A
 \mathchardef\PUpsilon="7107
 \mathchardef\PPi="7105
 \def\PB      {\ensuremath{B}\xspace}
 \def\PD      {\ensuremath{D}\xspace}
 \def\PJ      {\ensuremath{J}\xspace}
 \def\PK      {\ensuremath{K}\xspace}
 \def\Pb      {\ensuremath{b}\xspace}
 \def\Pc      {\ensuremath{c}\xspace}

 \def\Pp      {\ensuremath{p}\xspace}

 \def\Ps      {\ensuremath{s}\xspace}

 \def\thebaroffset{0.18em}
}
\newcommand{\offsetoverline}[2][\thebaroffset]{\kern #1\overline{\kern -#1 #2}}%

\makeatletter
\ifcase \@ptsize \relax
  \newcommand{\miniscule}{\@setfontsize\miniscule{4}{5}}
\or
  \newcommand{\miniscule}{\@setfontsize\miniscule{5}{6}}
\or
  \newcommand{\miniscule}{\@setfontsize\miniscule{5}{6}}
\fi
\makeatother

\DeclareRobustCommand{\optbar}[1]{\shortstack{{\miniscule (\rule[.5ex]{1.25em}{.18mm})}
  \\ [-.7ex] $#1$}}

\def\squark    {{\ensuremath{\Ps}}\xspace}

\def\cquark    {{\ensuremath{\Pc}}\xspace}
\def\cquarkbar {{\ensuremath{\overline \cquark}}\xspace}

\def\bquark    {{\ensuremath{\Pb}}\xspace}

\def\KorKbar {\kern \thebaroffset\optbar{\kern -\thebaroffset \PK}{}\xspace}

\def\D       {{\ensuremath{\PD}}\xspace}

\def\DorDbar {\kern \thebaroffset\optbar{\kern -\thebaroffset \PD}\xspace}

\def\Dp      {{\ensuremath{\D^+}}\xspace}
\def\Dm      {{\ensuremath{\D^-}}\xspace}

\def\DpDm    {\ensuremath{\Dp {\kern -0.16em \Dm}}\xspace}

\def\B       {{\ensuremath{\PB}}\xspace}

\def\BorBbar {\kern \thebaroffset\optbar{\kern -\thebaroffset \PB}\xspace}

\def\Bd      {{\ensuremath{\B^0}}\xspace}

\def\BdorBdbar {\kern \thebaroffset\optbar{\kern -\thebaroffset \Bd}\xspace}

\def\Bs      {{\ensuremath{\B^0_\squark}}\xspace}

\def\BsorBsbar {\kern \thebaroffset\optbar{\kern -\thebaroffset \Bs}\xspace}

\def\jpsi     {{\ensuremath{{\PJ\mskip -3mu/\mskip -2mu\Ppsi}}}\xspace}
\def\psitwos  {{\ensuremath{\Ppsi{(2S)}}}\xspace}

\def\Y#1S{\ensuremath{\PUpsilon{(#1S)}}\xspace}
\def\OneS  {{\Y1S}\xspace}

\def\ThreeS{{\Y3S}\xspace}

\def\proton      {{\ensuremath{\Pp}}\xspace}

\def\LorLbar     {\kern \thebaroffset\optbar{\kern -\thebaroffset \PLambda}\xspace}

\def\BF         {{\ensuremath{\mathcal{B}}}\xspace}
\def\BR         {\BF}

\def\AT#1     {\ensuremath{A_{\mathrm{T}}^{#1}}\xspace}           

\def\C#1      {\ensuremath{\mathcal{C}_{#1}}\xspace}                       
\def\Cp#1     {\ensuremath{\mathcal{C}_{#1}^{'}}\xspace}                    
\def\Ceff#1   {\ensuremath{\mathcal{C}_{#1}^{\mathrm{(eff)}}}\xspace}        
\def\Cpeff#1  {\ensuremath{\mathcal{C}_{#1}^{'\mathrm{(eff)}}}\xspace}       
\def\Ope#1    {\ensuremath{\mathcal{O}_{#1}}\xspace}                       
\def\Opep#1   {\ensuremath{\mathcal{O}_{#1}^{'}}\xspace}                    

\newcommand{\nospaceunit}[1]{\ensuremath{\text{#1}}}
\newcommand{\aunit}[1]{\ensuremath{\text{\,#1}}}

\newcommand{\tev}{\aunit{Te\kern -0.1em V}\xspace}
\newcommand{\gev}{\aunit{Ge\kern -0.1em V}\xspace}
\newcommand{\mev}{\aunit{Me\kern -0.1em V}\xspace}
\newcommand{\kev}{\aunit{ke\kern -0.1em V}\xspace}
\newcommand{\ev}{\aunit{e\kern -0.1em V}\xspace}

\newcommand{\mevc}{\ensuremath{\aunit{Me\kern -0.1em V\!/}c}\xspace}
\newcommand{\gevc}{\ensuremath{\aunit{Ge\kern -0.1em V\!/}c}\xspace}
\newcommand{\mevcc}{\ensuremath{\aunit{Me\kern -0.1em V\!/}c^2}\xspace}
\newcommand{\gevcc}{\ensuremath{\aunit{Ge\kern -0.1em V\!/}c^2}\xspace}

\def\mm   {\aunit{mm}\xspace}

\def\mum  {\ensuremath{\,\upmu\nospaceunit{m}}\xspace}

\def\nb {\aunit{nb}\xspace}
\def\invnb {\ensuremath{\nb^{-1}}\xspace}

\def\ps   {\ensuremath{\aunit{ps}}\xspace}

\def\gsim{{~\raise.15em\hbox{$>$}\kern-.85em
          \lower.35em\hbox{$\sim$}~}\xspace}
\def\lsim{{~\raise.15em\hbox{$<$}\kern-.85em
          \lower.35em\hbox{$\sim$}~}\xspace}

\def\sPlot{\mbox{\em sPlot}\xspace}

\def\sqsnn {\ensuremath{\protect\sqrt{s_{\scriptscriptstyle\text{NN}}}}\xspace}
\def\pt         {\ensuremath{p_{\mathrm{T}}}\xspace}

\def\ptot       {\ensuremath{p}\xspace}

\def\evtgen     {\mbox{\textsc{EvtGen}}\xspace}

\def\geant      {\mbox{\textsc{Geant4}}\xspace}

\def\photos     {\mbox{\textsc{Photos}}\xspace}

\def\pythia     {\mbox{\textsc{Pythia}}\xspace}

\def\tell1  {TELL1\xspace}
\def\ukl1   {UKL1\xspace}

\newcommand{\lhcborcid}[1]{\href{https://orcid.org/#1}{\hspace*{0.1em}\raisebox{-0.45ex}{\includegraphics[width=1em]{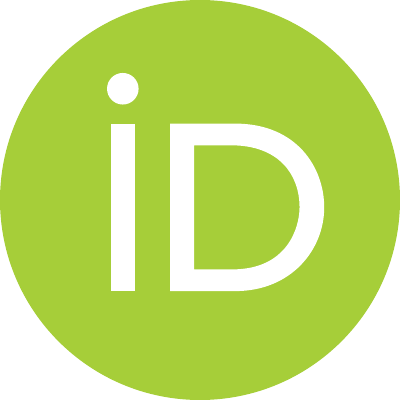}}}}

\usepackage{cite} 
\usepackage{mciteplus}

\usepackage{longtable} 
\usepackage{multirow}
\usepackage{float}
\begin{document}
\renewcommand{\thefootnote}{\fnsymbol{footnote}}
\setcounter{footnote}{1}
\begin{titlepage}
\pagenumbering{roman}

\vspace*{-1.5cm}
\centerline{\large EUROPEAN ORGANIZATION FOR NUCLEAR RESEARCH (CERN)}
\vspace*{1.5cm}
\noindent
\begin{tabular*}{\linewidth}{lc@{\extracolsep{\fill}}r@{\extracolsep{0pt}}}
\ifthenelse{\boolean{pdflatex}}
{\vspace*{-1.5cm}\mbox{\!\!\!\includegraphics[width=.14\textwidth]{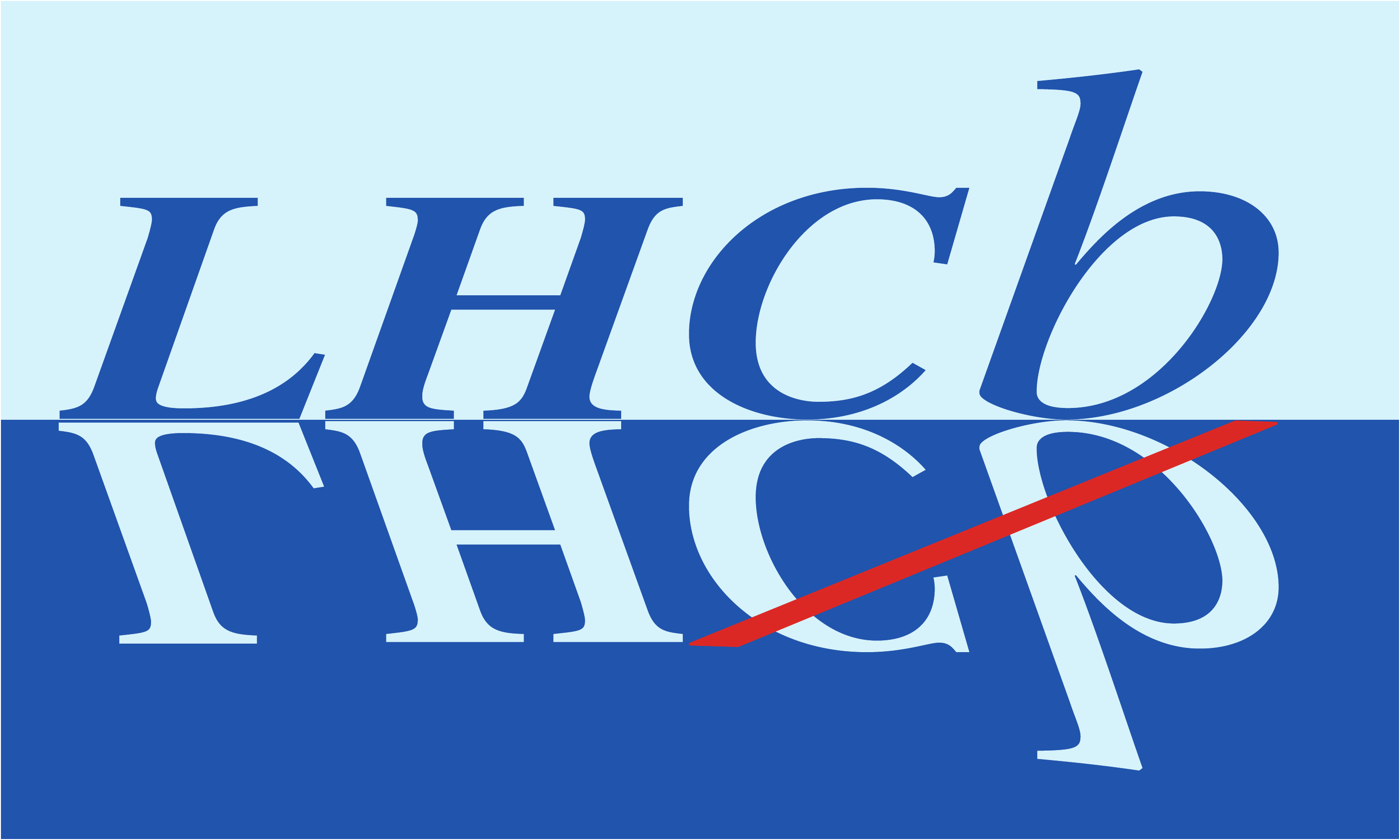}} & &}
 
\\
 & & CERN-EP-2025-114 \\  
 & & LHCb-PAPER-2025-011 \\  
 & & Feb 18, 2026 \\
 & & \\
\end{tabular*}

\vspace*{2.5cm}
{\normalfont\bfseries\boldmath\huge
\begin{center}
  \papertitle 
\end{center}
}

\vspace*{1.0cm}
\begin{center}
\paperauthors\footnote{Authors are listed at the end of this paper.}
\end{center}

\vspace{\fill}

\begin{abstract}
  \noindent
    The production ratio of \psitwos to \jpsi charmonium states is presented as a function of multiplicity in proton-lead collisions at a centre-of-mass energy of $\sqsnn=8.16$\tev, for both prompt and nonprompt sources. The total luminosity recorded by the \lhcb experiment corresponds to 13.6\invnb for $p$Pb collisions and 20.8\invnb for Pb$p$ collisions, where the first particle corresponds to the particle traveling towards the detector. Measurements are performed in the dimuon final state at forward (backward) centre-of-mass rapidity, with respect to the proton direction, $1.5<y^*<4.0$ ($-5.0<y^*<-2.5$)  for $p$Pb (Pb$p$) collisions.
    A multiplicity dependence of the prompt production ratio is observed in $p$Pb collisions, whereas no dependence is found in nonprompt production, nor in either prompt or nonprompt production in Pb$p$ collisions. These results suggest that in the Pb-going direction additional suppression mechanisms beyond comover effects may be present, possibly related to the formation of quark-gluon plasma. This highlights a transition from small to large collision systems and provides important insight into the suppression of charmonia in proton-nucleus collisions.
\end{abstract}

\vspace*{0.5cm}

\begin{center}
  Published in JHEP 11 (2025) 169
\end{center}
\vspace{\fill}
{\footnotesize 
\centerline{\copyright~\papercopyright. \href{\paperlicenceurl}{\paperlicence}.}}
\vspace*{2mm}
\end{titlepage}
\newpage
\setcounter{page}{2}
\mbox{~}

\renewcommand{\thefootnote}{\arabic{footnote}}
\setcounter{footnote}{0}
\cleardoublepage
\pagestyle{plain}
\setcounter{page}{1}
\pagenumbering{arabic}


\def\npv        {N^{\rm PV}_{\rm tracks}}
\def\nfwd       {N^{\rm PV}_{\rm fwd}}
\def\nbwd       {N^{\rm PV}_{\rm bwd}}
\def\nch        {\mathrm{d} N_{\rm ch} / \mathrm{d} \eta}
\def\nchf        {\frac{\mathrm{d} N_{\rm ch}} / {\mathrm{d} \eta}}

\section{Introduction} \label{sec:Introduction}

Heavy-quarkonium production has long been an essential probe for understanding Quantum Chromodynamics and its interactions in high-temperature environments. In particular, the study of quarkonium states, such as charmonia and bottomonia, in heavy-ion collisions provides valuable insights into the behaviour of strongly interacting matter under extreme conditions. Heavy-ion collisions at high energies are expected to create a hot and dense medium known as quark-gluon plasma (QGP), where quarks and gluons are deconfined. The production rate of heavy quarkonium states in these collisions is strongly influenced by the presence of QGP, being suppressed by  mechanisms such as Debye screening~\cite{Matsui:1986dk} or enhanced by quarkonium regeneration~\cite{Zhao:2020jqu}.

In proton-proton (\proton\proton) collisions, where the formation of QGP is not expected, quarkonium production can be described within the standard factorization framework~\cite{Brambilla:2010cs}, using proton parton distribution functions. However, in proton-nucleus~($p$A) collisions, such as proton-lead~($p$Pb), a $\cquark\cquarkbar$ pair may interact with nucleons while passing through the nucleus, leading to phenomena like parton energy loss due to multiple scattering~\cite{Arleo:2014oha}. Other effects such as gluon saturation and nuclear modifications of parton distribution functions may also affect quarkonium production~\cite{AtashbarTehrani:2017mzi}.

In this context, the study of excited and ground-state quarkonium production in $p$Pb collisions offers a unique opportunity to explore nuclear matter effects. Previous measurements of quarkonium production in $p$A collisions have demonstrated a stronger suppression of excited states, such as \ThreeS and \psitwos, with respect to their ground states, \OneS~\cite{LHCb-PAPER-2018-035} and \jpsi~\cite{LHCb-PAPER-2023-024,PHENIX:2022nrm,ALICE:2020vjy,LHCB-PAPER-2015-058,CMS:2025oqe,NuSea:1999mrl,NA50:2006rdp,PHENIX:2016vmz} respectively. This suppression is attributed to final-state effects, such as interactions with comoving particles~\cite{Capella:1996va,Ferreiro:2012rq}, where excited states, more loosely bound, are more susceptible to dissociation. An improved color evaporation model within the framework of the color glass condensate effective theory may also provide a partial explanation~\cite{Ma:2017rsu}. Recent studies have also explored the potential role of QGP formation in $p$A collisions~\cite{Du:2018wsj}, particularly through the suppression of the \psitwos state. 

This makes \psitwos a promising probe for investigating QGP-like effects in small collision systems. The possible formation of a small, short-lived QGP droplet in high-multiplicity small collision systems could lead to additional suppression mechanisms beyond cold nuclear matter effects~\cite{Du:2018wsj}.

Additionally, quarkonia production measurements in $p$Pb collisions serve as a valuable reference for future studies in larger systems, where the creation of QGP is expected to have significant impact. By establishing a baseline of nuclear matter effects in small systems, an improved understanding of the dynamics governing quarkonium production across different collision systems can be attained.

This study focuses on the production of \psitwos and \jpsi mesons in $p$Pb~(Pb$p$) collisions at a centre-of-mass energy of $\sqsnn = 8.16\tev$, using data collected in 2016 by the \lhcb experiment corresponding to an integrated luminosity of 13.6\invnb (20.8\invnb).

While the initial-state effects cancel in this ratio, final-state effects such as the relative contributions of nuclear effects and comoving particle interactions leading to quarkonium suppression can be studied. The production of both prompt and nonprompt \psitwos and \jpsi mesons is analysed, thereby separating the contributions of quarkonium states directly produced in hard scattering~(prompt) from those originating in $b$-hadron decays~(nonprompt). In both cases, the \psitwos to \jpsi production ratio is measured as a function of the charged-particle multiplicity in the event, which serves as a proxy for the amount of nuclear matter produced in the collision. In this way, the evolution of quarkonium production under different baryon density is probed. This aids in understanding the role of comover effects and enables investigation into whether the suppression mechanisms observed in heavy-ion collisions can also be detected in smaller systems like $p$Pb collisions.

\section{Detector, data and simulation}
\label{Data and Monte Carlo samples}

The \lhcb detector~\cite{LHCb-DP-2014-002} is a single-arm forward spectrometer covering the pseudorapidity range $2<\eta <5$, designed for the study of particles containing \bquark or \cquark quarks. The detector includes a high-precision tracking system consisting of a silicon-strip vertex detector surrounding the interaction region~(\velo), a large-area silicon-strip detector located upstream of a dipole magnet with a bending power of about $4{\mathrm{\,T\,m}}$, and three stations of silicon-strip detectors and straw drift tubes placed downstream of the magnet. The tracking system provides a measurement of the momentum, \ptot, of charged particles with a relative uncertainty that varies from 0.5\% at low momentum to 1.0\% at 200\gevc.
The minimum distance of a track to a primary collision vertex (PV), the impact parameter, is measured with a resolution of $(15+29/\pt)\mum$, where \pt is the component of the momentum transverse to the beam, in\,\gevc.

Muons, the \jpsi and \psitwos decay products, are identified by a system composed of alternating layers of iron and multiwire proportional chambers~\cite{LHCb-DP-2012-002}.

The data are collected under two configurations of the beams, $\proton$Pb and Pb$\proton$, with the first particle denoting the beam travelling in the forward direction $+z$, from the \velo towards the muon chambers. A higher baryon density is achieved in the Pb-going direction, which facilitates the investigation of charmonium suppression in a different final-state environment.
Simulated samples of \jpsi and \psitwos decays are required to model the effects of the detector acceptance and the imposed selection requirements. In the simulation, $\jpsi$ and $\psitwos$ signal events are generated using \pythia~\cite{Sjostrand:2007gs} with a specific \lhcb configuration~\cite{LHCb-PROC-2010-056}, and are embedded into inelastic $\proton$Pb and Pb$\proton$ collisions simulated with \textsc{Epos}~\cite{Pierog:2013ria} to ensure a realistic track multiplicity.

Decays of unstable particles are described by \evtgen~\cite{Lange:2001uf}, in which final-state radiation is generated using \photos~\cite{davidson2015photos}. The interaction of the generated particles with the detector, and its response, are implemented using the \geant toolkit~\cite{Allison:2006ve, *Agostinelli:2002hh} as described in Ref.~\cite{LHCb-PROC-2011-006}.

\section{Event selection}
\label{Selections}
In this analysis, the \jpsi and \psitwos mesons are reconstructed in their dimuon decay channels. The online event selection is performed by a trigger~\cite{LHCb-DP-2012-004}, which consists of a hardware stage, based on information from the calorimeter and muon systems, followed by a software stage, which performs a full event reconstruction. The trigger requires the transverse momentum of muons to be larger than 300~\mevc and the number of clusters reconstructed in the \velo detector to be less than 8000. The selection requirement on the \velo clusters removes less than $0.1\%$ of the high-multiplicity events. The software trigger further requires the tracks to 
be consistent with the muon hypothesis.

Offline selection criteria are subsequently applied to further discriminate against background. Global criteria on the events containing \jpsi and \psitwos candidates include restricting the number of PVs to exactly one to avoid multiple $\proton$Pb collisions occurring within the same beam interaction. Each muon-track candidate must satisfy $\pt >0.9$\gevc,  $2<\eta<5$ and a probability of being reconstructed from spurious hits~\cite{DeCian:2255039} smaller than 40\%. Particle identification~(PID) information is also employed to enhance consistency with the muon hypothesis. 
Pairs of oppositely charged tracks are required to originate from a common vertex and combined to form \jpsi and \psitwos candidates. Dimuon candidates must have masses within $\pm 120\mevcc$ of the known \jpsi or \psitwos mass~\cite{PDG2024}, $0.3 <\pt<14.0\gevc$, and rapidity in the centre-of-mass frame of the collisions of $1.5 < y^* < 4.0$ ($-5.0 < y^* < -2.5$) for $p$Pb (Pb$p$) collisions.

The pseudodecay time $t_z$, defined as 
\begin{equation}
    \label{PseudoDecayTime}
    t_z = \frac{(z_{X}-z_{\mathrm{PV}})  m_X}{p_z},
\end{equation}
is required to be $|t_z|<10\ps$ and its uncertainty to satisfy \mbox{$\sigma_{t_z}<0.3\ps$}. 
Furthermore, $X$ represents the $\jpsi$ or $\psitwos$ meson, $m_{X}$ its known mass, 
$z_{X}$ is the $z$ coordinate of its decay vertex and $z_{\rm PV}$ is the $z$-position of the associated PV. The pseudodecay time is used to distinguish between prompt and nonprompt production. Prompt mesons produced directly in the hard interaction typically have $t_z\sim0$, while nonprompt mesons from $b$-hadron decays exhibit broader $t_z$ distributions due to the $b$-hadron lifetime. 

To describe the charged-particle multiplicity close to the PV, three quantities, $\npv$, $\nfwd$, and $\nbwd$, are determined. Here, $\npv$ represents the number of tracks used for the PV reconstruction, while $\nfwd$ and $\nbwd$ represent its subset in the forward and backward pseudorapidity regions, respectively. 
As the charged-track acceptance is not uniform along $z_{\rm PV}$, which may induce biases in the track-multiplicity measurement, only the ranges in Table~\ref{TablePVZ} are considered for the analysis.

\begin{table}[!b]
\caption{Ranges of $z_{\rm PV}$ considered in this analysis,  where the indicated multiplicity variables are sufficiently uniform.}
\begin{center}
\renewcommand{\arraystretch}{1.1}
\begin{tabular}{llr}
\hline
Beam configuration & Multiplicity variable & \textbf{$z_{\rm PV}$}\\ 
\hline
        $p$Pb & $N^{\rm PV}_{\rm tracks}$ & $[-30, 180]\mm$\\
        $p$Pb & $N^{\rm PV}_{\rm fwd}$ & $[-180, 180]\mm$\\
        $p$Pb & $N^{\rm PV}_{\rm bwd}$ & $[-30, 180]\mm$\\
        Pb$p$ & $N^{\rm PV}_{\rm tracks}$ & $[-60, 180]\mm$\\
        Pb$p$ & $N^{\rm PV}_{\rm fwd}$ & $[-180, 120]\mm$\\
        Pb$p$ & $N^{\rm PV}_{\rm bwd}$ & $[-30, 180]\mm$\\
\hline
\end{tabular}
\end{center}
\label{TablePVZ}
\end{table}

The variable $\nch$, the density of charged primary tracks per unit of pseudorapidity, is also employed to allow comparison with other collision systems and experiments. The value of $\nch$ is derived from $\nfwd$ via a relationship established in simulation. For each $\nfwd$ bin, the uncertainty on the mean $\nch$ is obtained by combining the statistical effect due to the size of the simulation sample with the spread arising from alternative functional choices for the $\nfwd$ versus $\nch$ relation.

\section{Cross-section observables}
\label{RCS}
\def\effTot{\ensuremath{\epsilon_{\mathrm{tot}}}\xspace}
\def\effTotJ{\ensuremath{\epsilon_{\mathrm{tot,\jpsi}}}\xspace}
\def\effTotP{\ensuremath{\epsilon_{\mathrm{tot,\psitwos}}}\xspace}
\def\effAcc{\ensuremath{\epsilon_{\mathrm{acc}}}\xspace}
\def\effReco{\ensuremath{\epsilon_{\mathrm{Reco\&Sel}}}\xspace}
\def\effID{\ensuremath{\epsilon_{\mathrm{MuonID}}}\xspace}
\def\effTrigger{\ensuremath{\epsilon_{\mathrm{Trigger}}}\xspace}

The cross-section for charmonia production is given by
\begin{equation}
    \sigma
    = \frac{1}{\BR\mathcal{L}}
    \frac{N}
           {\effTot}=\frac{Y}{\BR\mathcal{L}} \,,
  \label{CrossSecJ}
\end{equation}
where $\BR$ is the charmonia branching fraction to the dimuon channel, $\mathcal{L}$ is the integrated luminosity and $Y$ is the measured charmonia yield $N$, corrected by the total efficiency \effTot. 
The cross-section ratio is therefore,
\begin{equation}
 \frac{\sigma^i_{\psitwos}}{\sigma^i_{\jpsi}}  =
    \frac{\mathcal{B}_{\jpsi \rightarrow \mu^+ \mu^-}}{\mathcal{B}_{\psitwos\rightarrow \mu^+ \mu^-}}\frac{Y_{\psitwos}^i}{Y_{\jpsi}^i}\,,
    \label{Rsingle}
\end{equation}
for a given multiplicity bin $i$. The intervals for each multiplicity variable are chosen so that the number of \psitwos candidates in each bin is roughly equal. 
The multiplicity bins are composed from the three multiplicity variables described earlier, divided by their respective mean values in no-bias data $\langle\npv\rangle_{\rm NB}$, $\langle\nfwd\rangle_{\rm NB}$, and $\langle\nbwd\rangle_{\rm NB}$. As this sample is recorded based on the \lhc beam clock, with no trigger or requirements other than a bunch-bunch crossing~\cite{LHCb-DP-2012-004}, their mean multiplicity values provide an unbiased reference. The cross-section ratio in each multiplicity bin is further normalised by the ratio of the integrated cross-sections as
\begin{equation}
 \mathrm{Normalised \ }\frac{\sigma^i_{\psitwos}}{\sigma^i_{\jpsi}}  = 
\frac{Y^i_{\psitwos} / \sum_jY^j_{\psitwos}}{Y^i_{\jpsi}/\sum_jY^j_{\jpsi}}\,,
    \label{Rratio}
\end{equation}
where the \psitwos and \jpsi branching fractions cancel in the double ratio.

Alternatively, to compare the cross-section ratio across different collision systems, the branching fractions for \jpsi and \psitwos decays, and the respective uncertainties, are removed by multiplying both sides of Eq.~\ref{Rsingle} as
\begin{equation}
\frac{\BR_{\psitwos}\sigma^i_{\psitwos}}{\BR_{\jpsi}\sigma^i_{\jpsi}} = 
    \frac{Y^i_{\psitwos}}{Y^i_{\jpsi}}\,.
    \label{Nratio}
\end{equation}

\section{Signal extraction and efficiency correction}
\label{SE}
The yields of prompt and nonprompt charmonia are determined through an extended two-dimensional unbinned maximum-likelihood fit, performed on the mass $m_{\mu^+\mu^-}$ and pseudodecay time $t_z$ distributions of the charmonia candidates. The fit is performed separately in each multiplicity bin, with the results reported in Appendix~\ref{app:ratios}. Only  the intervals listed in Table~\ref{TablePVZ} are considered. 
The mass distribution of the \jpsi signal is modelled with a sum of two Crystal Ball~(CB) functions~\cite{Skwarnicki:1986xj} sharing a common peak position but with distinct widths. For \psitwos mesons, a single CB function is used. The peak positions and widths of the CB functions are allowed to vary in the fit, while all other parameters are fixed from simulation. The background contribution is described with an exponential function with a slope parameter varying in the fit.

\begin{figure}[!tbp]
  \begin{center}
    \includegraphics[width=0.5\linewidth]{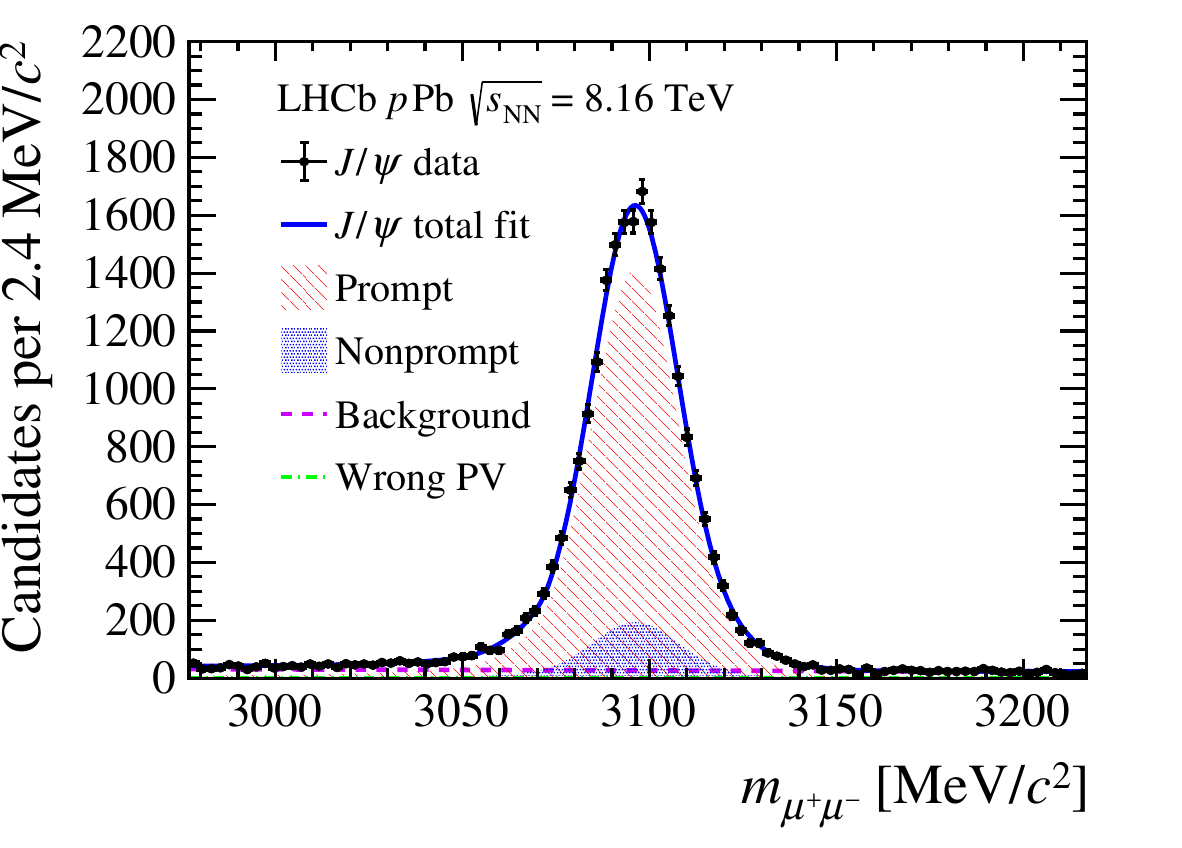}%
    \includegraphics[width=0.5\linewidth]{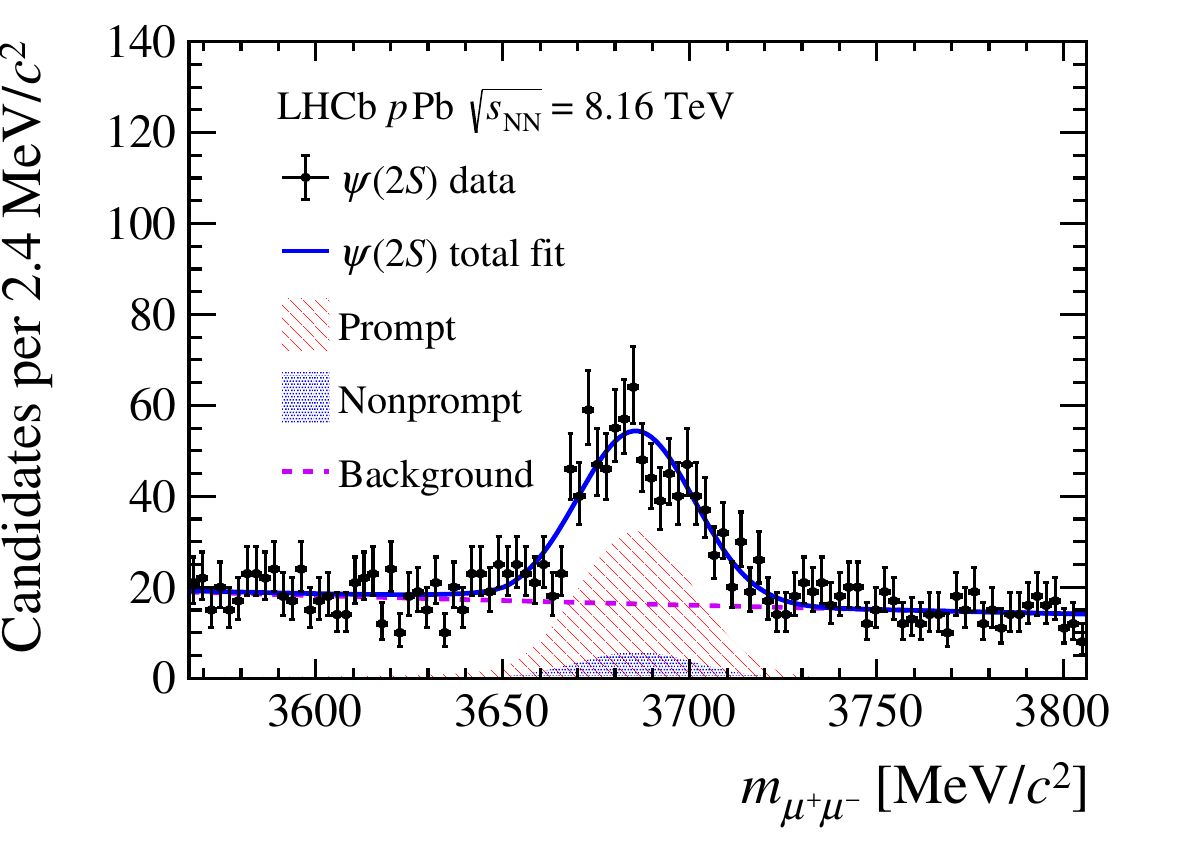}
    \includegraphics[width=0.5\linewidth]{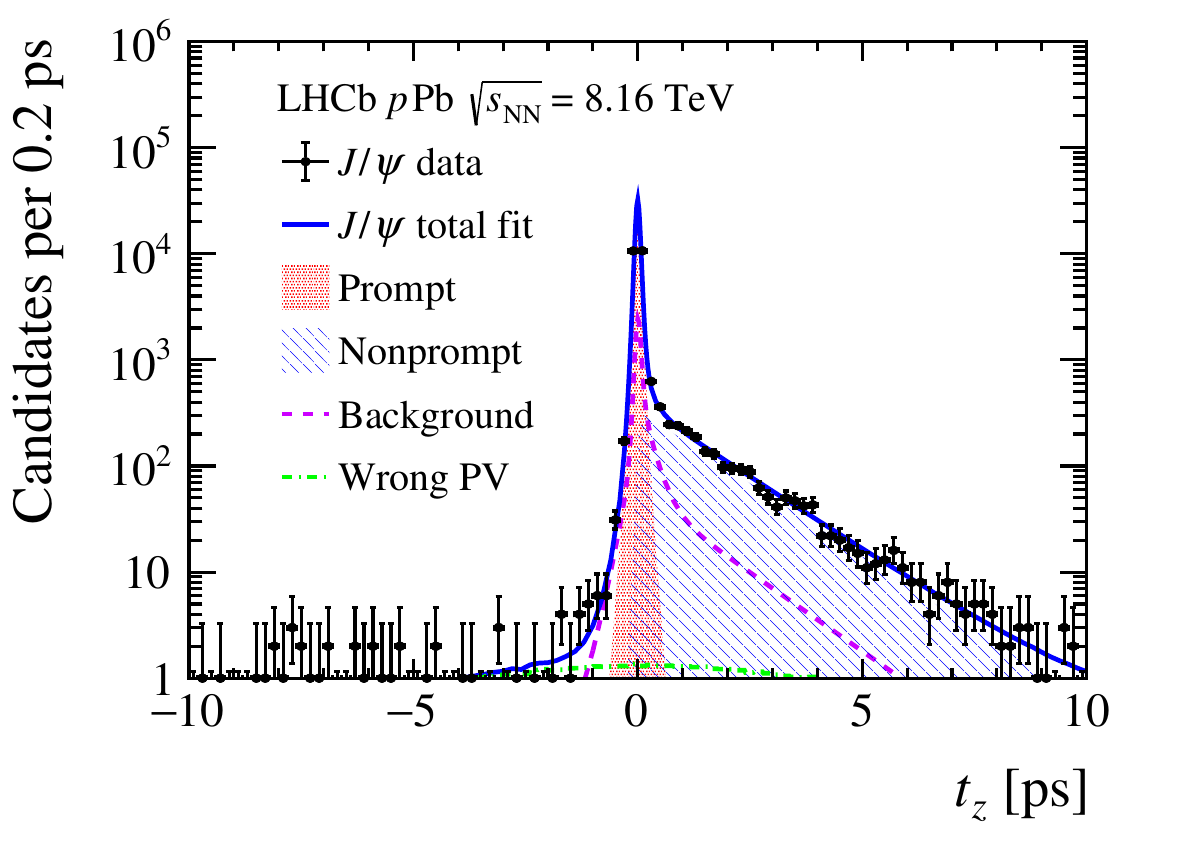}%
    \includegraphics[width=0.5\linewidth]{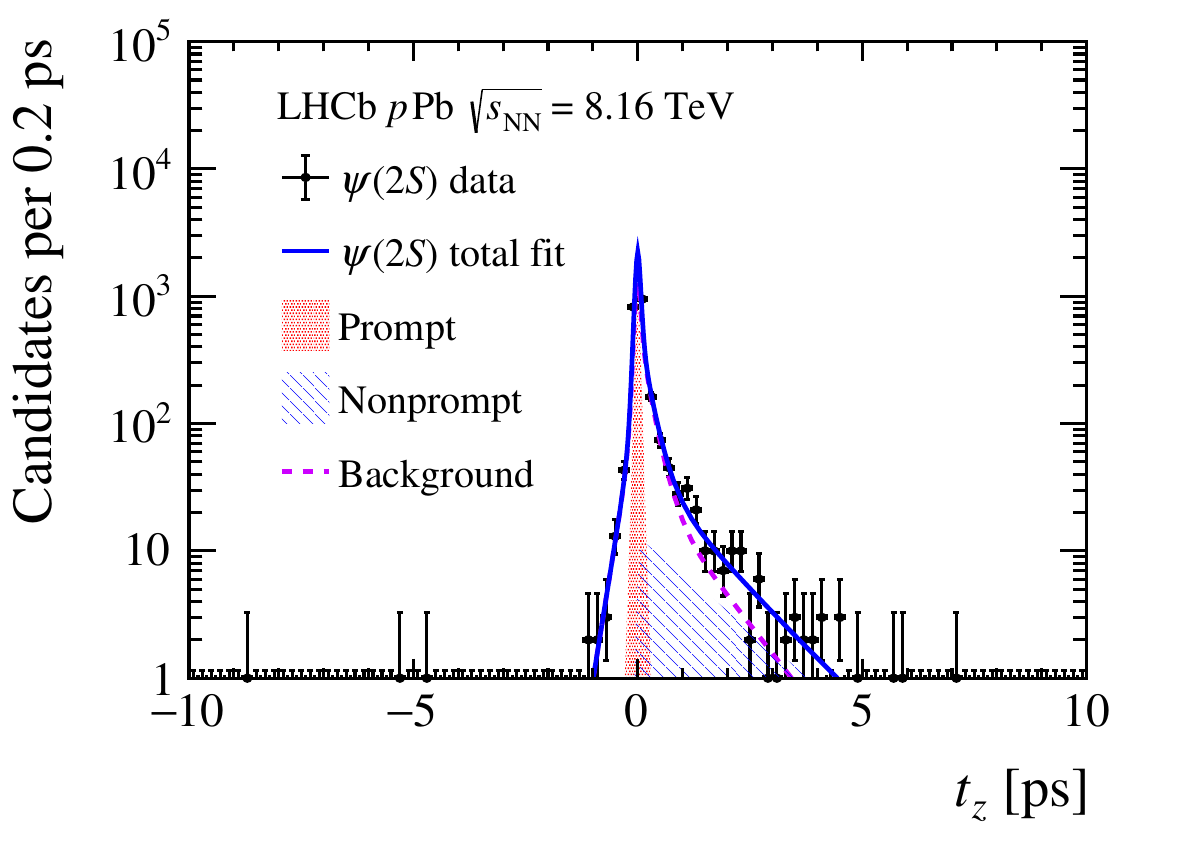}
  \end{center}
  \caption{
  Distributions of (top) ~$m_{\mu^+\mu^-}$ and (bottom) ~$t_z$ for (left) \jpsi and (right) \psitwos candidates in the interval $4 \leq \npv <$45, $1.5<y^*<4.0$, and $0.3<\pt<14.0\gevc$ in the $p$Pb configuration. The fit results are also shown. 
    }
  \label{fig:2Dtz}
\end{figure}

For the $t_z$ distributions, prompt and nonprompt charmonia are modelled with a Dirac $\delta$ function and an exponential-decay function defined from only positive $t_z$, respectively, both convolved with a shared resolution function described by the sum of two Gaussian functions with common mean and different widths. The mean is treated as a free parameter in the fit, while the ratio between the widths of the two Gaussian functions is fixed from simulation. The combinatorial background $t_z$ distribution is parametrised empirically with the sum of three exponential functions defined in the positive $t_z$ domain and the sum of two exponential functions for the negative domain, each convolved with the same resolution function. The parameters for the background shape are determined from fits to the data in the mass-sideband regions $60<|m_{\mu^+\mu^-}-m_{\jpsi}|<120\mevcc$ and $60<|m_{\mu^+\mu^-}-m_{\psitwos}|<120\mevcc$, where mostly background candidates are present. These shape parameters are then fixed for the baseline two-dimensional fit. For the $\jpsi$ candidates, the $t_z$ distribution also includes a tail component to account for the residual presence of candidates associated with the wrong PV, in cases where the true PV is not reconstructed and a nearby PV is incorrectly associated with the signal. This shape is extracted by deliberately assigning each event with the PV from a separate event, and used as a template in the two-dimensional fit. Given a negligible amount of wrong-PV candidates in the \psitwos sample, a tail component is not considered there. An example two-dimensional fit result projected onto the $m_{\mu^+\mu^-}$ and $t_z$ dimensions is presented in Fig.~\ref{fig:2Dtz}. 

The fit results are also used to weight the $\pt$ and $y^*$ distributions in simulation to match those of signal-weighted data obtained with the \sPlot technique~\cite{Pivk:2004ty}. The generator-level simulation sample is also weighted to provide a more accurate determination of the detector acceptance. After the weighting, the \jpsi and \psitwos yields are corrected by the total efficiency, which includes several components: acceptance efficiency, defined as the fraction of \jpsi or \psitwos decays where both muons lie within the \lhcb acceptance ($2 < \eta < 5$); reconstruction and selection efficiency, which accounts for the probability that both muons are reconstructed as tracks and pass the selection criteria; PID efficiency, defined as the probability that both muons satisfy the particle identification requirements; and trigger efficiency, which reflects the probability that the event passes the trigger selections. The total efficiency ratio between \jpsi and \psitwos mesons is expressed as the product of the corresponding efficiency ratios for each component. The track-reconstruction and muon-identification efficiencies are determined using the data-driven methods described in Ref.~\cite{LHCb-PUB-2016-021,LHCb-DP-2013-002}. The trigger efficiency is estimated from simulation and cross-checked using data~\cite{LHCb-DP-2012-004}.

\section{Systematic uncertainties}
\label{Systematic uncertainty}
Several sources of systematic uncertainty are identified.
The uncertainties arising from the limited size of the calibration data are taken into account, including the PID~\cite{LHCb-PUB-2016-021} and the tracking-efficiency correction ratios~\cite{LHCb-DP-2013-002}, as well as the $\pt$ and $y^*$ calibration using signal-weighted samples extracted from data by the \sPlot method. Residual imperfections in simulating the detector acceptance, accounted for through a similar weighting of $\pt$ and $y^*$, are considered separately for generator-level simulation. Uncertainties deriving from the simulated sample size are propagated to the final results using a large number of pseudoexperiments. An additional uncertainty related to the kinematic binning scheme chosen for the evaluation of the PID efficiencies is also considered as the variation on the cross-section values when using alternative binning schemes.

Uncertainty due to the choice of the mass fit model is studied by replacing the single CB function for \psitwos by the sum of two CB functions with a common mass peak position, as used for the baseline \jpsi model. The scale factor between the two width parameters, and the remaining tail parameters, are determined from simulation. The difference from the baseline result is found to be negligible and is therefore not considered further. The uncertainty arising from imperfections in the $t_z$ signal model is assessed by comparing the fitted yields of prompt and nonprompt candidates in simulation with their corresponding true values. The observed differences are negligible compared to the statistical uncertainty, and thus no additional systematic uncertainty is assigned.

The trigger efficiency estimated from simulation in the baseline approach is compared to that obtained from a fully reconstructed data sample, following the data-driven method described in Ref.~\cite{LHCb-DP-2012-004}. The resulting shift with respect to the baseline approach is assigned as a systematic uncertainty.

As the online selection on the \velo clusters removes less than $0.1\%$ of the events, this value is conservatively taken as a systematic uncertainty. 
All uncertainties are summarised in Table~\ref{AllSysErr} and found to be subdominant with respect to the statistical contribution.

\begin{table}[tb]
\caption{Relative statistical and systematic uncertainty ($\%$) ranges on the normalised ratio of \psitwos to \jpsi cross-sections in different $\npv$ intervals.}
\begin{center}
\begin{tabular}{lrr}
\hline
Source & $p$Pb & Pb$p$ \\
\hline

        Tracking, PID, \pt and
        $y^*$ calibration & 1.7--3.6 & 2.1--3.6 \\

        Kinematic acceptance & 0.8--1.2 & 0.9--1.3 \\

        Simulated sample size & $<0.1$ & $<0.1$ \\

        PID kinematic binning scheme & 0.4--1.7 & 0.1--1.8 \\

        Fit model & $<0.1$ & $<0.1$\\

        Trigger efficiency & 3.2--3.9 & 3.6--4.1 \\

        VELO clusters selection & 0.1 & 0.1 \\

        Total & 4.0--5.5 & 4.2--5.5 \\
\hline
Statistical & 5.8--15.5&  5.2--13.3 \\ \hline
\end{tabular}
\label{AllSysErr}
\end{center}
\end{table}


\section{Results}
\label{Results}

\begin{figure}[tb]
  \begin{center}
    \includegraphics[width=0.5\linewidth]{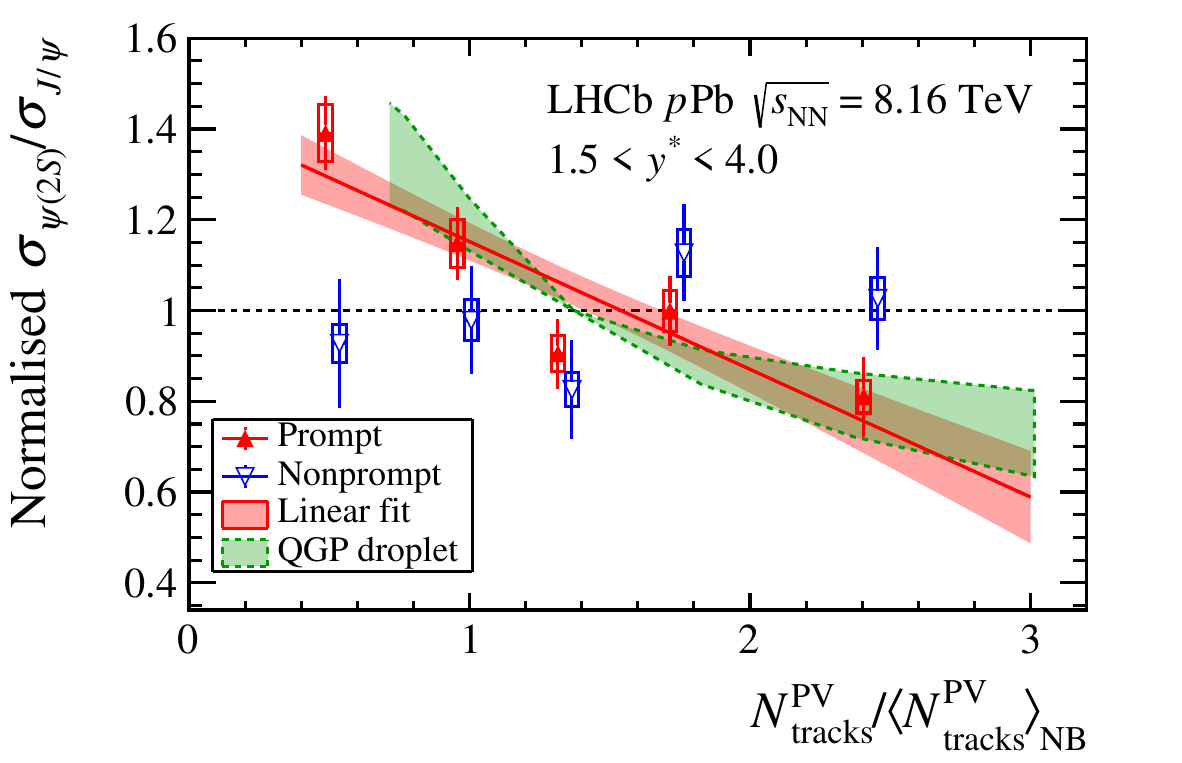}%
    \includegraphics[width=0.5\linewidth]{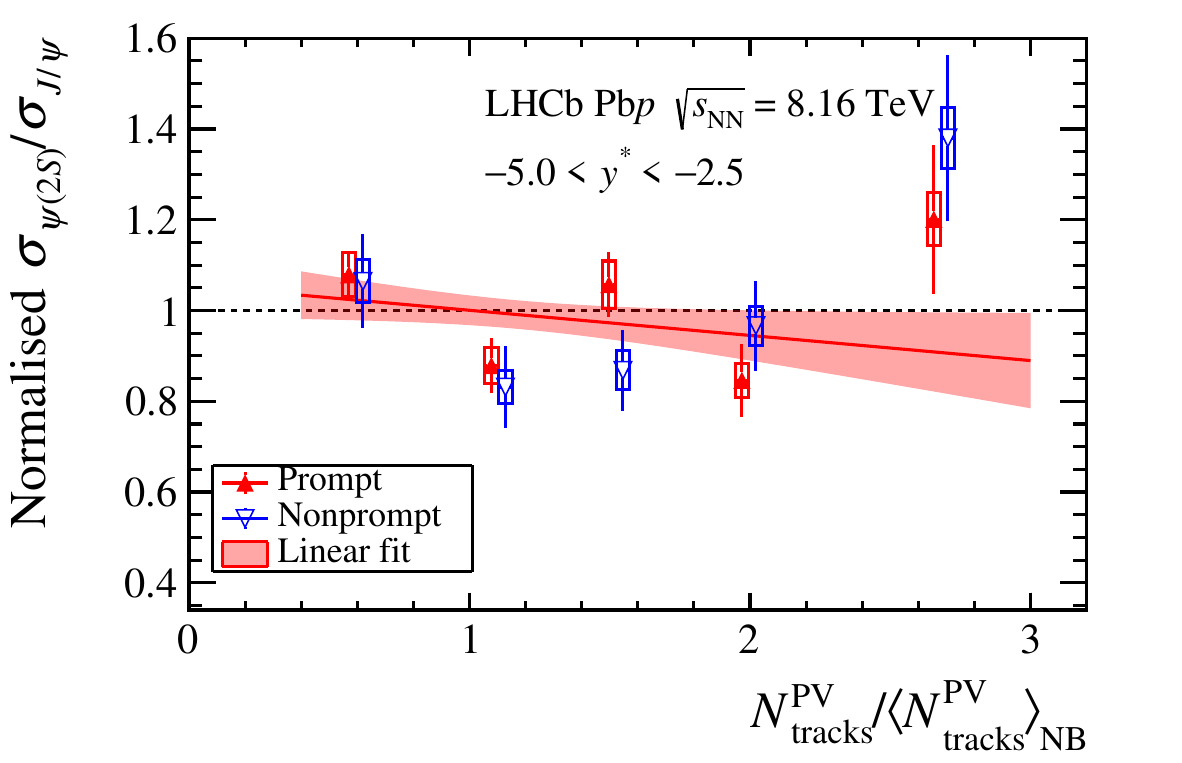}
  \end{center}
  \caption{Normalised ratio of \psitwos to \jpsi cross-sections as a function of $\npv/\langle\npv\rangle_{\rm NB}$ for the (left) $p$Pb and (right) Pb$p$ beam configurations. The error bars and height of each box represent statistical and systematic uncertainties, respectively. The red band stands for a linear fit for prompt ratio with confidence level $68\%$ and the green band in the left is theoretical prediction~\cite{Wen:2022utn,Bai:2024xmm}. The $x$-axis values for nonprompt ratios are slightly offset to the right to improve readability.
    }
  \label{RPVN}
\end{figure}

Figure~\ref{RPVN} shows the normalised ratio of \psitwos to \jpsi cross-sections as a function of $\npv/\langle\npv\rangle_{\rm NB}$. 
The nonprompt production ratio exhibits no significant dependence on $\npv$ in either beam configuration. This behaviour is expected, since variations in $b$-hadron production across multiplicity bins affect the absolute cross-sections of the \psitwos and \jpsi mesons individually, but largely cancel in their ratio. 

In contrast, for the $p$Pb configuration, the prompt production ratio shows a clear decreasing trend with $\npv$ in the $p$Pb configuration.  Specifically, the hypothesis of a constant normalised ratio is strongly disfavored, with the p-value of $ 5\times10^{-6}$. A linear fit $f(x) = p_1 ( x-x_0) + p_0$, where $x_0$ is the mean of $x$, provides a good description of the observed decrease, as shown in Fig.~\ref{RatioPVN}, yielding $p_1=-0.281\pm0.057$ and $p_0=1.046\pm 0.036$ with a chisquare probability of $P_L(\chi^2)=0.10$. This multiplicity dependence is consistent with theoretical prediction of charmonium production modification due to hot-medium effects from QGP droplets~\cite{Wen:2022utn,Bai:2024xmm}. However, for the Pb$p$ configuration, the data are consistent with the absence of a significant dependence on multiplicity, where the linear fit result gives $p_1=-0.055\pm0.054$ and $p_0=0.970\pm0.038$ with $P_L(\chi^2)=0.01$, and the fit with a constant gives a p-value of $0.04$.

\begin{figure}[tb]
  \begin{center}
    \includegraphics[width=0.5\linewidth]{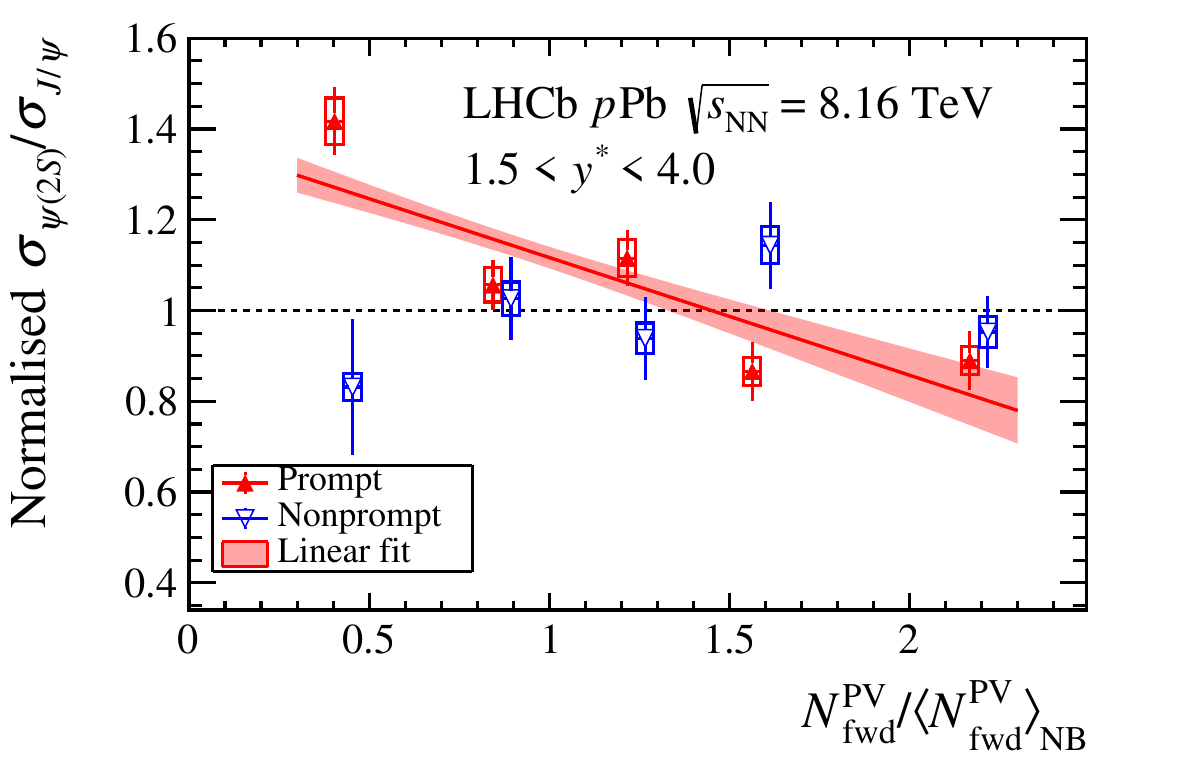}%
    \includegraphics[width=0.5\linewidth]{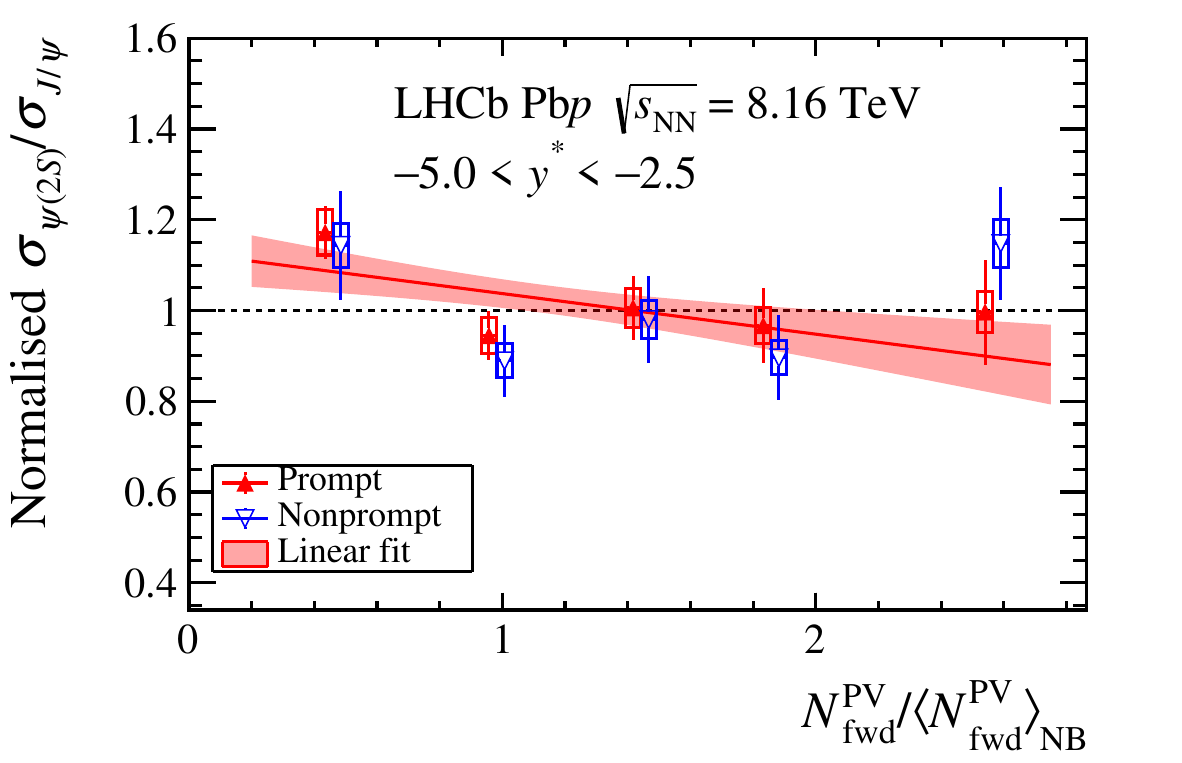}
    \vspace*{-0.5cm}
  \end{center}
  \caption{Normalised ratio of \psitwos to \jpsi cross-sections as a function of $\nfwd/\langle\nfwd\rangle_{\rm NB}$ for the (left) $p$Pb and (right) Pb$p$ beam configurations. The error bars and height of each box represent statistical and systematic uncertainties, respectively. The red band stands for a linear fit for the prompt ratio with confidence level $68\%$. The $x$-axis values for nonprompt ratios are slightly offset to the right to improve readability.
    }
  \label{RFor}
\end{figure}

The dependence on $\nfwd/\langle\nfwd\rangle_{\rm NB}$, measured in the same rapidity region as the charmonia, is shown in Fig.~\ref{RFor}. A pattern consistent with that observed for $\npv$ is found. Nonprompt production shows no significant trend, while prompt production displays a clear suppression of the \psitwos relative to the \jpsi in the $p$-going direction, but not in the Pb-going direction. In the $p$Pb configuration, a linear fit gives $p_1=-0.259\pm 0.050$ and $p_0=1.055\pm 0.028$ with $P_L(\chi^2)=0.01$, while the fit of a constant gives a p-value of $5\times 10^{-8}$. The linear fit to  the Pb-going configuration data gives $p_1=-0.089\pm 0.051$ and $p_0=0.998\pm0.034$ with $P_L(\chi^2)=0.12$, while the fit with a constant has p-value of 0.08. In the $p$Pb configuration the linear fit might not be describing the data well, but it still out-performed the fit with a constant and the slope shows a negative trend with more than $5\sigma$. A pronounced forward-backward asymmetry is evident: the result in the \proton-going direction is consistent with observations in \proton\proton collisions~\cite{LHCb-PAPER-2023-035}, while the Pb-going result shows no significant multiplicity dependence. 

\begin{figure}[tb]
  \begin{center}
    \includegraphics[width=0.5\linewidth]{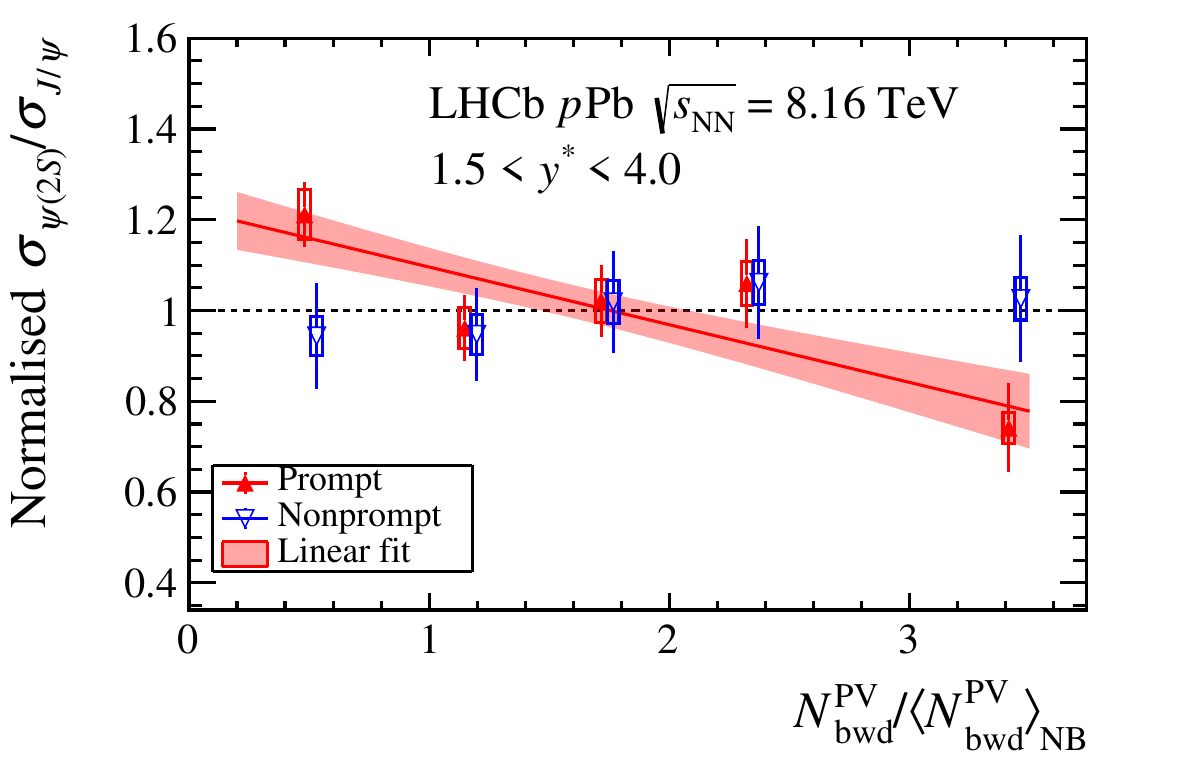}%
    \includegraphics[width=0.5\linewidth]{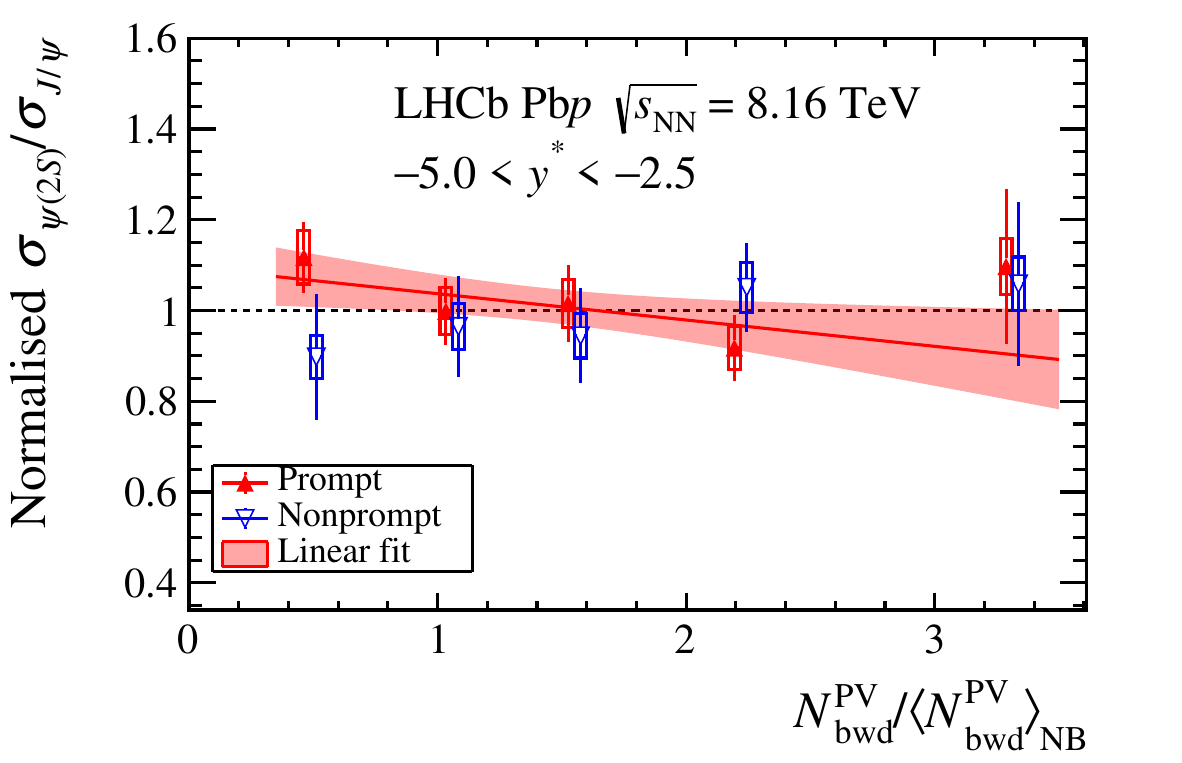}
    \vspace*{-0.5cm}
  \end{center}
  \caption{Normalised ratio of \psitwos to \jpsi cross sections as a function of $\nbwd/\langle\nbwd\rangle_{\rm NB}$ for the $p$Pb~(left) and Pb$p$~(right) beam configurations. The error bars and height of each box represent statistical and systematic uncertainties, respectively. The red band stands for a linear fit for the prompt ratio with confidence level $68\%$. The $x$-axis values for nonprompt ratios are slightly offset to the right to improve readability.
    }
  \label{RBack}
\end{figure}

The ratio as a function of $\nbwd/\langle\nbwd\rangle_{\rm NB}$ is presented in Fig.~\ref{RBack}. Here, $\nbwd$ is measured in the rapidity region opposite to that of the charmonia, thereby reducing the direct correlation between the multiplicity observable and the charmonium production. In the $p$Pb configuration, the decreasing trend of the normalised ratio observed with $\npv$ and $\nfwd$ is still present, but less pronounced, with a fit of $p_1=-0.127\pm 0.038$ and $p_0=0.992\pm0.038$ with $P_L(\chi^2)=0.17$ and a fit of constant with p-value 0.005. This indicates that when multiplicity is measured away from the charmonium rapidity, the comover contribution can be disentangled, leading to a flatter dependence compared to the ratio as function of $\nfwd/\langle\nfwd\rangle_{\rm NB}$ and $\npv/\langle\npv\rangle_{\rm NB}$. In the Pb$p$ configuration, the fit yields $p_1=-0.058\pm0.050$ and $p_0=0.996\pm0.040$ and fit of a constant with p-value of 0.56, which is fully compatible with the constant hypothesis.

The absence of \psitwos suppression with multiplicity in Pb$p$ collisions, exhibiting a distinct behaviour from $pp$ and $p$Pb collisions but consistent with PbPb observations, suggests similar physical effects may occur in Pb$p$ and PbPb systems, but not in $pp$ or $p$Pb, potentially indicating QGP-like conditions in Pb-going configurations despite the smaller system size.

\begin{figure}[tb]
  \begin{center}
  \includegraphics[width=0.7\linewidth]{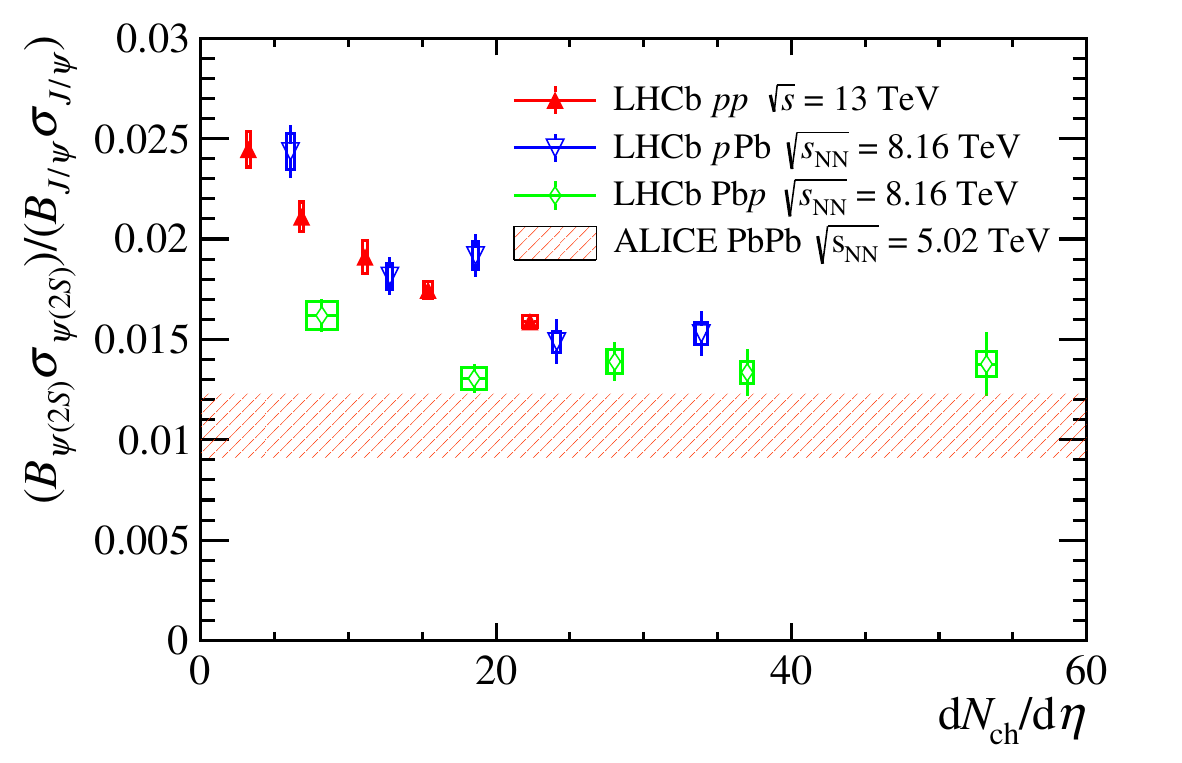}
  \end{center}
  \caption{Comparisons of the ratio of the \psitwos to \jpsi cross-sections in $pp$, $p$Pb, Pb$p$ and PbPb collisions~\cite{ALICE:2022jeh}. The error bars and height of each box represent statistical and systematic uncertainties, respectively, and the width of the box represents the uncertainty on the mean value of $\mathrm{d}N_{\rm ch}/{\mathrm{d}\eta}$ within each bin.}
\label{pppPb}
\end{figure}

The multiplicity dependence of the ratio of \psitwos to \jpsi cross sections scaled by the branching fractions according to Eq.~\ref{Nratio} is compared in Fig.~\ref{pppPb} between different collision systems.
The multiplicity dependence observed in $p$Pb collisions is consistent with the results from the $pp$ configuration, indicating a similar environment in the final states of both collision systems. However, in Pb$p$ collisions, the achieved higher baryon density results in an environment more similar to PbPb collisions and the observed decreasing trend is more similar to the result from PbPb collisions measured by \alice~\cite{ALICE:2022jeh}. This suggests that $p$Pb and $pp$ collisions might be affected by comovers as a dominant suppression mechanism, while additional suppression mechanisms beyond comovers, such as QGP effects, could exist for the Pb\textit{p} case. This additional suppression mechanism, which distinguishes Pb$p$ from $p$Pb, represents a transition toward large collision systems and is important for understanding the underlying dynamics in these systems.

\section{Conclusion}
\label{conclusion}
The ratio of \psitwos to \jpsi cross-sections at a centre-of-mass energy \mbox{$\sqsnn=8.16\tev$} is measured with a data sample collected by the \lhcb experiment in 2016 corresponding to an integrated luminosity of $13.6\pm0.3\invnb$ for $p$Pb collisions and $20.8\pm0.5\invnb$ for Pb$p$ collisions. 
The normalised ratio of prompt and nonprompt \psitwos to \jpsi cross-sections, as functions of different multiplicity metrics, are measured for $0.3<\pt<14.0\gevc$ and $1.5<y^*<4.0$ ($-5.0<y^*<-2.5$) for $p$Pb~(Pb$p$) collisions.
In $p$Pb collisions, a decreasing trend is seen in the ratio of prompt production as a function of $\npv$ and $\nfwd$, with the decrease as a function of $\nbwd$ being less pronounced. Conversely, for Pb$p$ collisions, which have a higher mean multiplicity compared to $p$Pb collisions, the prompt and nonprompt ratios do not show a decreasing trend with multiplicity. The overall ratios of the efficiency-corrected charmonium yields in $p$Pb and Pb$p$ collisions are compared with measurements from other collision systems at different energies. A forward-backward asymmetry is observed, where in the $p$-going direction the result is compatible with that seen in $pp$ collisions. The suppression patterns in Pb$p$ collisions cannot be fully explained by comovers alone. The larger suppression of \psitwos suggests an additional mechanism, possibly linked to QGP-like effects. The forward-backward asymmetry indicates a transition between small  and large collision systems, making Pb$p$ collisions a unique probe for QGP-like phenomena in small systems.

\section*{Acknowledgements}
%
%
\noindent We express our gratitude to our colleagues in the CERN
accelerator departments for the excellent performance of the LHC. We
thank the technical and administrative staff at the LHCb
institutes.
We acknowledge support from CERN and from the national agencies:
ARC (Australia);
CAPES, CNPq, FAPERJ and FINEP (Brazil); 
MOST and NSFC (China); 
CNRS/IN2P3 (France); 
BMBF, DFG and MPG (Germany); 
INFN (Italy); 
NWO (Netherlands); 
MNiSW and NCN (Poland); 
MCID/IFA (Romania); 
MICIU and AEI (Spain);
SNSF and SER (Switzerland); 
NASU (Ukraine); 
STFC (United Kingdom); 
DOE NP and NSF (USA).
We acknowledge the computing resources that are provided by ARDC (Australia), 
CBPF (Brazil),
CERN, 
IHEP and LZU (China),
IN2P3 (France), 
KIT and DESY (Germany), 
INFN (Italy), 
SURF (Netherlands),
Polish WLCG (Poland),
IFIN-HH (Romania), 
PIC (Spain), CSCS (Switzerland), 
and GridPP (United Kingdom).
We are indebted to the communities behind the multiple open-source
software packages on which we depend.
Individual groups or members have received support from
Key Research Program of Frontier Sciences of CAS, CAS PIFI, CAS CCEPP, 
Fundamental Research Funds for the Central Universities,  and Sci.\ \& Tech.\ Program of Guangzhou (China);
Minciencias (Colombia);
EPLANET, Marie Sk\l{}odowska-Curie Actions, ERC and NextGenerationEU (European Union);
A*MIDEX, ANR, IPhU and Labex P2IO, and R\'{e}gion Auvergne-Rh\^{o}ne-Alpes (France);
Alexander-von-Humboldt Foundation (Germany);
ICSC (Italy); 
Severo Ochoa and Mar\'ia de Maeztu Units of Excellence, GVA, XuntaGal, GENCAT, InTalent-Inditex and Prog.~Atracci\'on Talento CM (Spain);
SRC (Sweden);
the Leverhulme Trust, the Royal Society and UKRI (United Kingdom).

\clearpage
\section*{Appendices}

\appendix

\addcontentsline{toc}{section}{Appendices}

\section{Table of ratios}
\label{app:ratios}

The prompt and nonprompt \psitwos to \jpsi cross-section ratios multiplied by their branching ratios in different multiplicity bins are listed in Tables~\ref{RatioPVN}--\ref{RatioBack}. 
The total ratio, which enters Eq.~\ref{Rratio} and is used to produce Figs.~\ref{RPVN}--\ref{RBack}, acts as a normalisation factor to scale the prompt and nonprompt ratios around unity. This scaling factor only results in a relative vertical shift and therefore the uncertainties on the total ratios are not quoted.

\begin{table}[H]
\caption{Ratio (\%) of production cross-section for \psitwos to \jpsi. The first uncertainties are statistical, the second are the systematic uncertainties.}
\centering
\resizebox{\linewidth}{!}{
\begin{tabular}{ c c c | c c c }
\hline
\multicolumn{3}{ c }{$p$Pb} &\multicolumn{3}{ c }{Pb$p$}\\
\hline
$N_{\rm tracks}^{\rm PV}$ & Prompt & Nonprompt & $N_{\rm tracks}^{\rm PV}$ & Prompt & Nonprompt \\
\hline
4--45 & $0.172\pm0.010\pm0.008$ & $0.218\pm0.034\pm0.012$ & 4--60 & $0.115\pm0.006\pm0.005$ & $0.243\pm0.024\pm0.011$ \\
45--70 & $0.142\pm0.010\pm0.007$ & $0.230\pm0.028\pm0.011$ & 60--90 & $0.094\pm0.007\pm0.004$ & $0.190\pm0.021\pm0.009$ \\
70--90 & $0.112\pm0.010\pm0.005$ & $0.194\pm0.026\pm0.009$ & 90--120 & $0.113\pm0.008\pm0.006$ & $0.199\pm0.020\pm0.011$ \\
90--120 & $0.124\pm0.010\pm0.006$ & $0.265\pm0.025\pm0.011$ & 120--160 & $0.090\pm0.009\pm0.004$ & $0.221\pm0.023\pm0.009$ \\
120--270 & $0.100\pm0.011\pm0.004$ & $0.241\pm0.027\pm0.010$ & 160--330 & $0.128\pm0.017\pm0.006$ & $0.316\pm0.042\pm0.014$ \\
\hline
4--270  & 0.123 & 0.233 & 4--330 & 0.107 & 0.229 \\
\hline
\end{tabular}
}
\label{RatioPVN}
\end{table}

\begin{table}[H]
\caption{Ratio (\%) of production cross-section for \psitwos to \jpsi. The first uncertainties are statistical, the second are the systematic uncertainties.}
\centering
\resizebox{\linewidth}{!}{
\begin{tabular}{ c c c | c c c }
\hline
\multicolumn{3}{ c }{$p$Pb} & \multicolumn{3}{ c }{Pb$p$}\\
\hline
$N_{\rm fwd}^{\rm PV}$ & Prompt & Nonprompt & $N_{\rm fwd}^{\rm PV}$ & Prompt & Nonprompt \\
\hline
0--25 & $0.183\pm0.010\pm0.007$ & $0.192\pm0.034\pm0.009$ & 0--35 & $0.122\pm0.006\pm0.005$ & $0.264\pm0.028\pm0.011$ \\
25--43 & $0.136\pm0.007\pm0.005$ & $0.237\pm0.021\pm0.009$ & 35--65 & $0.098\pm0.006\pm0.004$ & $0.205\pm0.018\pm0.009$ \\
43--57 & $0.144\pm0.008\pm0.005$ & $0.217\pm0.021\pm0.008$ & 65--85 & $0.104\pm0.007\pm0.004$ & $0.226\pm0.022\pm0.010$ \\
57--72 & $0.112\pm0.008\pm0.004$ & $0.264\pm0.022\pm0.008$ & 85--110 & $0.100\pm0.009\pm0.004$ & $0.207\pm0.022\pm0.009$ \\
72--150 & $0.115\pm0.008\pm0.004$ & $0.220\pm0.018\pm0.008$ & 110--250 & $0.103\pm0.012\pm0.005$ & $0.265\pm0.029\pm0.011$ \\
\hline
0--150 & 0.129 & 0.230 & 0--250 & 0.103 & 0.231 \\
\hline
\end{tabular}
}
\label{RatioFor}
\end{table}

\begin{table}[H]
\caption{Ratio (\%) of production cross-section for \psitwos to \jpsi. The first uncertainties are statistical, the second are the systematic uncertainties.}
\centering
\resizebox{\linewidth}{!}{
\begin{tabular}{ c c c | c c c }
\hline
\multicolumn{3}{ c }{$p$Pb} & \multicolumn{3}{ c }{Pb$p$}\\
\hline
$N_{\rm bwd}^{\rm PV}$ & Prompt & Nonprompt & $N_{\rm bwd}^{\rm PV}$ & Prompt & Nonprompt \\
\hline
0--17 & $0.151\pm0.009\pm0.007$ & $0.219\pm0.027\pm0.011$ & 0--13 & $0.116\pm0.008\pm0.006$ & $0.205\pm0.032\pm0.012$ \\
17--29 & $0.120\pm0.009\pm0.006$ & $0.220\pm0.024\pm0.010$ & 13--22 & $0.104\pm0.008\pm0.005$ & $0.220\pm0.025\pm0.012$ \\
29--40 & $0.127\pm0.010\pm0.006$ & $0.237\pm0.026\pm0.011$ & 22--30 & $0.105\pm0.009\pm0.005$ & $0.216\pm0.024\pm0.012$ \\
40--54 & $0.132\pm0.012\pm0.006$ & $0.247\pm0.029\pm0.011$ & 30--47 & $0.095\pm0.008\pm0.005$ & $0.240\pm0.022\pm0.012$ \\
54--180 & $0.092\pm0.012\pm0.004$ & $0.238\pm0.033\pm0.009$ & 47--120 & $0.114\pm0.018\pm0.006$ & $0.242\pm0.041\pm0.014$ \\
\hline
0--180 & 0.125 & 0.232 & 0--120 & 0.104 & 0.228 \\
\hline
\end{tabular}
}
\label{RatioBack}
\end{table}

\clearpage
\addcontentsline{toc}{section}{References}
\bibliographystyle{LHCb}
\bibliography{main,standard,LHCb-PAPER,LHCb-CONF,LHCb-DP,LHCb-TDR}
\newpage
\centerline
{\large\bf LHCb collaboration}
\begin
{flushleft}
\small
R.~Aaij$^{38}$\lhcborcid{0000-0003-0533-1952},
A.S.W.~Abdelmotteleb$^{57}$\lhcborcid{0000-0001-7905-0542},
C.~Abellan~Beteta$^{51}$\lhcborcid{0009-0009-0869-6798},
F.~Abudin{\'e}n$^{57}$\lhcborcid{0000-0002-6737-3528},
T.~Ackernley$^{61}$\lhcborcid{0000-0002-5951-3498},
A. A. ~Adefisoye$^{69}$\lhcborcid{0000-0003-2448-1550},
B.~Adeva$^{47}$\lhcborcid{0000-0001-9756-3712},
M.~Adinolfi$^{55}$\lhcborcid{0000-0002-1326-1264},
P.~Adlarson$^{84}$\lhcborcid{0000-0001-6280-3851},
C.~Agapopoulou$^{14}$\lhcborcid{0000-0002-2368-0147}
% ...

{\footnotesize \it

% ...
$^{14}$Universit{\'e} Paris-Saclay, CNRS/IN2P3, IJCLab, Orsay, France\\
% ...
$^{38}$Nikhef National Institute for Subatomic Physics, Amsterdam, Netherlands\\
% ...
$^{47}$Instituto Galego de F{\'\i}sica de Altas Enerx{\'\i}as (IGFAE), Universidade de Santiago de Compostela, Santiago de Compostela, Spain\\
% ...
$^{51}$Physik-Institut, Universit{\"a}t Z{\"u}rich, Z{\"u}rich, Switzerland\\
% ...
$^{55}$H.H. Wills Physics Laboratory, University of Bristol, Bristol, United Kingdom\\
% ...
$^{57}$Department of Physics, University of Warwick, Coventry, United Kingdom\\
% ...
$^{60}$School of Physics and Astronomy, University of Glasgow, Glasgow, United Kingdom\\
$^{61}$Oliver Lodge Laboratory, University of Liverpool, Liverpool, United Kingdom\\
% ...
$^{69}$Syracuse University, Syracuse, NY, United States\\
% ...
$^{84}$Department of Physics and Astronomy, Uppsala University, Uppsala, Sweden, associated to $^{60}$\\
% ...
}
\end{flushleft}
\end{document}